\documentstyle[amsfonts,aps]{revtex}

\newtheorem{theorem}{Theorem}

\newtheorem{definition}[theorem]{Definition}

\newtheorem{remark}[theorem]{Remark}

\begin{document}
\title{PRESERVING ENTANGLEMENT UNDER DECOHERENCE AND SANDWICHING ALL SEPARABLE
STATES}
\author{Robert B. Lockhart$^{1}$ and Michael J. Steiner$^{2}$}
\address{1 Mathematics Department United States Naval Academy, Annapolis, Maryland\\
21402, e-mail rbl@usna.edu; 2 Naval Research Laboratory, Washington\\
D.C. e-mail mjs@mike.nrl.navy.mil}
\date{\today }
\maketitle

\begin{abstract}
Every entangled state can be perturbed, for instance by decoherence, and
stay entangled. For a large class of pure entangled states, which includes
all bipartite and all maximally entangled ones, we show how large the
perturbation can be. Maximally entangled states can be perturbed the most.
For each entangled state in our class, we construct hyperplanes which
sandwich the set of all separable states. As the number of particles, or the
dimensions of the Hilbert spaces for two of the particles, increases, the
distance between two of these hyperplanes goes to zero.
\end{abstract}

\bigskip

\section{\protect\bigskip INTRODUCTION}

Quantum systems display many properties, which are not observed in the
macroscopic world. One of the most fascinating is entanglement. Starting
with the fundamental paper of Einstein, Podolsky, and Rosen \cite{EPR} and
continuing through the works of Bell \cite{Bell} and others to the present
day, it has played a key role in debates over the foundations, completeness,
and interpretation of quantum mechanics \cite{Bohm},\cite{Peres},\cite{Ormes}%
.

Today entanglement is playing a key role in the burgeoning field of quantum
information\cite{popescu},\cite{williams}. It is fundamental to
teleportation \cite{brassard}, \cite{vitali}{\bf ,}\cite{werner}, secure key
distribution \cite{shor}{\bf ,}\cite{losalamos}, dense coding \cite{werner},
quantum error correction,\cite{bennet},\cite{barnes} and other applications 
\cite{note}. Thus increasingly researchers around the world are working
with, or at least trying to work with, entangled states, both in the
laboratory and on the computer.

Obviously then, it is important to know how much one can perturb an
entangled state and still have entanglement. Such considerations arise when
one takes decoherence into account in applications and when one has
approximations and error (round off and otherwise) in computer simulations
or algorithms.

In this paper, for a large class of entangled pure states, we construct open
neighborhoods of pure and mixed entangled states. The class of entangled
states we do this for includes all bipartite, entangled, pure states, all
multiparticle maximally entangled states, and many others. In particular,
suppose we have $p$-particles with the $\alpha -th$ one modelled on ${\Bbb C}%
^{n_{\alpha }}$, $\alpha =1,...,p$. Then our system of $p$-particles is
modelled on the Hilbert space ${\Bbb C}^{N}={\Bbb C}^{n_{1}}\otimes \cdots
\otimes {\Bbb C}^{n_{p}}$. Let $\left\{ \mid \psi _{\alpha }^{j_{\alpha
}}\rangle \right\} $ be an orthonormal basis for ${\Bbb C}^{n_{\alpha }}$.
Then $\left\{ \mid \psi _{1}^{j_{1}}\cdots \psi _{p}^{j_{p}}\rangle \right\} 
$ is an orthonormal basis for ${\Bbb C}^{N}$. Without loss of generality,
assume $n_{1}\leq \cdots \leq n_{p}$. The entangled states we consider in
this paper are the ones of the form

\begin{equation}
\psi =\sum_{j=1}^{n_{1}}v_{j}\mid \psi _{1}^{j}\cdots \psi _{p}^{j}\rangle
\label{1}
\end{equation}
where $\sum_{j=1}^{n_{1}}\left| v_{j}\right| ^{2}=1$ and no $v_{j}$ equals
1. (If a $v_{j}=1,$ then $\psi $ would not be entangled.)

Note that by using a Schmidt decomposition, every bipartite state can be
expressed in this form. So too is every maximally entangled state on ${\Bbb C%
}^{n}\otimes \cdots \otimes {\Bbb C}^{n}$ of this form. Indeed they are the
ones with each $v_{j}=\frac{1}{\sqrt{n}}$. Finally, notice that by using
local operations, i.e. acting on ${\Bbb C}^{N}$ by $U(n_{1})\otimes \cdots
\otimes U(n_{p})$, any state $\varphi =\sum_{j_{1}=1}^{n_{1}}\cdots
\sum_{j_{p}=1}^{n_{p}}w_{j_{1}\cdots j_{p}}\mid \psi _{1}^{j_{1}}\cdots \psi
_{p}^{j_{p}}\rangle $ can be expressed in the form of (1), provided for each 
$j_{i}$ there is at most one $w_{j_{1}\cdots j_{p}}$ \ which is not 0.

The neighborhood, $G_{\psi }$, of entangled states we construct for $\psi $
lies in the set of all mixed and pure states, since this is the physically
reasonable thing to do. Thus we need to express $\psi $ in terms of its
density matrix, $E_{\psi }$. For each $E_{\psi }$, we find the distance to
the nearest, pure, product states. What we find is the following:

\begin{theorem}
Let $E_{\psi }$ be the entangled, pure state $\sum_{j,k=1}^{n_{1}}v_{j}%
\overline{v_{k}}\mid \psi _{1}^{j}\cdots \psi _{p}^{j}\rangle \langle \psi
_{1}^{k}\cdots \psi _{p}^{k}\mid $. The closest, pure, product states to $%
E_{\psi }$ are a distance $\sqrt{2(1-\left| v_{j_{0}}\right| ^{2})}$ away
from $E_{\psi }$, where $\left| v_{j_{o}}\right| =\max \left\{ \left|
v_{1}\right| ,...,\left| v_{p}\right| \right\} .$ An example of such a
closest, pure product state is the projection, $S_{\psi }$ =$\mid \psi
_{1}^{j_{0}}\cdots \psi _{p}^{j_{0}}\rangle \langle \psi _{1}^{j_{0}}\cdots
\psi _{p}^{j_{0}}\mid .$
\end{theorem}

We shall give the proof of this theorem in the next section. For now, let us
describe how it is used to construct $G_{\psi }$, which is quite simple.
Take \ $C_{\psi }$ to be the $N-1$ dimensional hyperplane which\ contains $%
S_{\psi }$ and is perpendicular to the line, $L_{\psi }$ , connecting $%
E_{\psi }$ with $\frac{1}{N}I$, the totally mixed state. Similarly, let $%
F_{\psi }$ be the parallel hyperplane, which contains any projection which
is furthest away from $E_{\psi }$. (These are precisely the projections
which commute with $E_{\psi }$ a separable example of which is $R=\mid \psi
_{1}^{1}\cdots \psi _{p-1}^{1}\psi _{p}^{2}\rangle \langle \psi
_{1}^{1}\cdots \psi _{p-1}^{1}\psi _{p}^{2}\mid .$) Then we have the
following theorem

\begin{theorem}
All separable states either lie on one of the hyperplanes $C_{\psi }$ or $%
F_{\psi \text{ }}$ or lie between them. Thus every state outside the
sandwich formed by $C_{\psi }$ and $F_{\psi }$ is entangled. This region, $%
G_{\psi },$ outside the $C_{\psi }$, $F_{\psi }$ sandwich is an open,
connected neighborhood of $E_{\psi }=$ $\sum_{j,k=1}^{n_{1}}v_{j}\overline{%
v_{k}}\mid \psi _{1}^{j}\cdots \psi _{p}^{j}\rangle \langle \psi
_{1}^{k}\cdots \psi _{p}^{k}\mid $. A state $Q$ is in $G_{\psi }$ if and
only if $\left\langle Q,E_{\psi }\right\rangle =Trace(QE_{\psi })>\left|
v_{j_{0}}\right| ^{2}$\ where, as before, $\left| v_{j_{0}}\right| =\max
\left\{ \left| v_{1}\right| ,...,\left| v_{p}\right| \right\} _{.}$
\end{theorem}

Again we shall postpone the proof until the next section and instead shall
now make a few remarks and give one last theorem.

\begin{remark}
$G_{\psi }$ will be largest when, $\left| v_{j_{0}}\right| ^{2}$ is
smallest. Since $\sum_{j=1}^{n_{1}}\left| v_{j}\right| ^{2}=1$, this occurs
when $\left| v_{1}\right| ^{2}=\cdots =\left| v_{p}\right| ^{2}=\frac{1}{%
n_{1}}$. Thus it is these states, which are the maximally entangled ones
when $n_{1}=\cdots =n_{p}$, that can withstand the greatest amount of
decoherence and still be in $G_{\psi }$ and so remain entangled.
\end{remark}

\begin{remark}
Since, the plane $C_{\psi }$ contains the separable state $S_{\psi }$, there
is no larger neighborhood of $E_{\psi }$, consisting solely of entangled
states, given by an inequality, $\left\langle Q,E_{\psi }\right\rangle >K,$
than $G_{\psi }$. In this sense $G_{\psi }$ is the largest neighborhood of $%
E_{\psi }$ consisting solely of entangled states. It may, however, not
contain the largest ball of entangled states centered at $E_{\psi }$.
\end{remark}

\begin{remark}
It is well known \cite{pit},\cite{braun},\cite{caves}\cite{dur} that if $%
E_{\psi }$ is a maximally entangled state on ${\Bbb C}^{n}\otimes \cdots
\otimes {\Bbb C}^{n}$, then the separable state on the line $L_{\psi }$,
which connects $E_{\psi }$ with $\frac{1}{N}I,$ that is closest to $E_{\psi
} $ is $W(s)=(1-s)\frac{1}{N}I+sE_{\psi }$, where $s=(1+n^{p-1})^{-1}$. When 
$p=2$, it is easy to compute that this state lies in the hyperplane $C_{\psi
} $. This has two important consequences: a) of all separable states, not
just those on $L_{\psi }$, the state $W(s)$ is the closest to $E_{\psi }$,
and b) the neighborhood, $G_{\psi }$, contains the largest open ball of
entangled states centered at $E_{\psi }$. Thus in this case $G_{\psi }$ is
the largest physically usable neighborhood of $E_{\psi }$ consisting solely
of entangled states. When $p>2$, the state $W(s)$ lies inside the sandwich
formed by $C_{\psi }$ and $F_{\psi }$. This means $G_{\psi }$ might not, in
this case, contain the largest ball of entangled states centered at $E_{\psi
}$. It also means that, in this case, $W(s)$ is not the closest separable
state to $E_{\psi }$. Indeed, simple geometry shows the line which contains $%
E_{\psi }$ and intersects the line connecting $W(s)$ with $S_{\psi }$
perpendicularly, intersects that line at a separable state which is closer
to $E_{\psi }$ than $W(s)$.
\end{remark}

\begin{remark}
From the last example given in the remark just made, it should be clear that
we do not claim all states between $C_{\psi }$ and $F_{\psi }$ are
separable. Many are entangled. In fact, numerical simulation for low
dimensional bipartite cases indicates that a large percentage of the states
inside the sandwich are entangled. However, there are no separable states
outside the sandwich.
\end{remark}

To finish this introduction, we shall state our last theorem. Basically it
says that for a system modelled on ${\Bbb C}^{n_{1}}\otimes \cdots \otimes 
{\Bbb C}^{n_{p}}$ with $n_{1}\leq \cdots \leq n_{p}$, the thickness of the
thinnest sandwich which contains all separable states goes to 0 as $n_{p-1}$
or $p$ increases to infinity. This means that for systems with a large
number of particles, or with at least two particles modelled on large
dimensional Hilbert spaces, all separable states cluster near a hyperplane
which contains the totally mixed state. Before stating the theorem we have
to make the following definition.

\begin{definition}
For the set of integers $\left\{ n_{1},...,n_{p}\right\} $, and any
partition $\pi $ of the set into two subsets, $\left\{ \left\{ n_{\pi
_{1}},...,n_{\pi _{k}}\right\} ,\left\{ n_{\pi _{k+1}},...,n_{\pi
_{p}}\right\} \right\} ,$ let $f(\pi )=\min (n_{\pi _{1}}\cdots n_{\pi
_{k}},n_{\pi _{k+1}}\cdots n_{\pi _{p}})$. Then $\kappa (n_{1},...,n_{p})$
is the maximum over all partitions of $\left\{ n_{1},...,n_{p}\right\} $
into two subsets of $1/f(\pi ).$
\end{definition}

For example, if the system consists of $p$-qubits, then $\kappa =2^{-m}$ if $%
p=2^{2m}$ and $\kappa =2^{-(m-1)}$ if $p=2^{2m-1\text{.}}.$ On the other
hand if, for instance, the system is modelled on ${\Bbb C}^{2}\otimes {\Bbb C%
}^{3}\otimes {\Bbb C}^{4}\otimes {\Bbb C}^{30}$, then $\kappa =\frac{1}{24}.$
In all cases, $\kappa \leq (n_{1}n_{3}\cdots n_{p-1})^{-1}$ if $p$ is odd
and $\kappa \leq \max ((n_{p-1}n_{1}n_{3}\cdots
n_{p-2})^{-1},(n_{p}n_{2}\cdots n_{p-3})^{-1})$ if $p$ is even.

\begin{theorem}
Consider a quantum system modelled on ${\Bbb C}^{n_{1}}\otimes \cdots
\otimes {\Bbb C}^{n_{p}}.$ There exist parallel hyperplanes which are a
distance $\kappa \sqrt{\frac{N}{N-1}}$ apart and which have the property
that all separable states either lie on one of the planes or lie between
them. In particular for every separable state $T$ the largest ball of
separable states centered at $T$ must have a radius no bigger than $\kappa 
\sqrt{\frac{N}{N-1}}$.
\end{theorem}

\section{PROOFS OF THEOREMS}

In this section we prove our theorems, starting with the first. For $\psi
=\sum_{j=1}^{n_{1}}v_{j}\mid \psi _{1}^{j}\cdots \psi _{p}^{j}\rangle $, the
associated projection is $E_{\psi }=\sum_{j,k=1}^{n_{1}}v_{j}\overline{v_{k}}%
\mid \psi _{1}^{j}\cdots \psi _{p}^{j}\rangle \langle \psi _{1}^{j}\cdots
\psi _{p}^{j}\mid $. For $\mu =1,...,p$, take $A_{\mu }$ to be the
projection $\sum_{j_{\mu },k_{\mu }=1}^{n_{\mu }}a_{\mu j_{\mu }}\overline{%
a_{\mu k_{\mu }}}\mid \psi _{\mu }^{j_{\mu }}\rangle \langle \psi _{\mu
}^{k_{\mu }}\mid $ on ${\Bbb C}^{n_{\mu }}$ and $A$ to be the pure, product
projection $A_{1}\otimes \cdots \otimes A_{p}$. Then $A=%
\sum_{j_{1},k_{1}=1}^{n_{1}}\cdots
\sum_{j_{p},k_{p}=1}^{n_{p}}a_{1j_{1}}\cdots a_{pj_{p}}\overline{a_{1k_{1}}}%
\cdots \overline{a_{pk_{p}}}\mid \psi _{1}^{j_{1}}\cdots \psi
_{p}^{j_{p}}\rangle \langle \psi _{1}^{j_{1}}\cdots \psi _{p}^{j_{p}}\mid $.

We want to find the smallest distance from such an $A$ to $E_{\psi }$. To do
so, first note the square of the distance from $A$ to $E_{\psi }$ is

\begin{equation}
\left\| E_{\psi }-A\right\| ^{2}=\left\langle E_{\psi }-A,E_{\psi
}-A\right\rangle =\left\langle E_{\psi },E_{\psi }\right\rangle -2%
\mathop{\rm Re}%
\left\langle E_{\psi },A\right\rangle +\left\langle A,A\right\rangle
\end{equation}
Since $E_{\psi }$ and $A$ are positive semi-definite Hermitian operators,
their inner product is real and equals $Tr(E_{\psi }A)$. Hence $\left\|
E_{\psi }-A\right\| ^{2}=2(1-Tr(E_{\psi }A)).$ This will be minimum when $%
Tr(E_{\psi }A)=\sum_{k=1}^{n_{1}}v_{k}\overline{a_{1k}}\cdots \overline{%
a_{pk}}\sum_{j=1}^{n_{1}}\overline{v_{j}}a_{1j}\cdots a_{pj}$ is maximum.

Setting $\Phi =\sum_{j=1}^{n_{1}}\overline{v_{j}}a_{1j}\cdots a_{pj}$, we
see that $Tr(E_{\psi }A)=\Phi \overline{\Phi }=\left| \Phi \right| ^{2}$. In
turn $\left| \Phi \right| ^{2}=\left| \sum_{j=1}^{n_{1}}\overline{v_{j}}%
a_{1j}\cdots a_{pj}\right| ^{2}\leq (\sum_{j=1}^{n_{1}}\left| v_{j}\right|
\left| a_{1j}\right| \cdots \left| a_{pj}\right|
)^{2}=(\sum_{j=1}^{n_{1}}\left| v_{j}\right| r_{1j}\cdots r_{pj})^{2}$,
where $r_{\mu j}=\left| a_{\mu j}\right| $. This last expression is
equivalent to $\left| \left\langle \rho ,V\beta \right\rangle \right| ^{2}$,
where $V$ is the $n_{1}\times n_{1}$ diagonal matrix with the $\left|
v_{j}\right| $ as the diagonal entries and $\rho $ and $\beta $ are the $%
n_{1}$ dimensional vectors with the components $\rho _{j}=r_{2j}\cdots
r_{pj} $ and $\beta _{j}=r_{1j}$. Using the Cauchy-Schwarz inequality and
the definition of the operator norm of a matrix, we obtain the inequality $%
\left| \left\langle \rho ,V\beta \right\rangle \right| ^{2}\leq \left\| \rho
\right\| ^{2}\left\| V\right\| _{op}^{2}\left\| \beta \right\| ^{2}$. Since $%
V$ is a diagonal matrix, $\left\| V\right\| _{op}^{2}=\max \left|
v_{j}\right| ^{2}=\left| v_{j_{0}}\right| ^{2}$. Furthermore, by assumption $%
\sum_{j=1}^{n_{1}}r_{\mu j}^{2}=1$ and so $\left\| \beta \right\| ^{2}=1$
and $\left\| \rho \right\| ^{2}\leq 1.$ Thus $Tr(E_{\psi }A)=\left|
\left\langle \rho ,V\beta \right\rangle \right| ^{2}\leq \left|
v_{j_{0}}\right| ^{2}$. Noting that if $S_{\psi }=$ $\mid \psi
_{1}^{j_{0}}\cdots \psi _{p}^{j_{0}}\rangle \langle \psi _{1}^{j_{1}}\cdots
\psi _{p}^{j_{p}}\mid $, then $Tr(E_{\psi }S_{\psi })=\left|
v_{j_{0}}\right| ^{2},$ we obtain the proof of the first theorem.

The proof of the second, basically, uses simple vector operations and facts
from trigonometry. For two states $K$ and $Q$, take $V(K,Q)$ to be the
vector with tail at $K$ and head at $Q$. As above, take $S_{\psi }$ to be
any of the closest pure product states to $E_{\psi }$ and consider the
triangle whose sides are $V(\frac{1}{N}I,E_{\psi }),$ $V(\frac{1}{N}%
I,S_{\psi }),$ and $V(S_{\psi },E_{\psi })$. Since $E_{\psi }$ and $S_{\psi
} $ are rank 1 projections, the length of the first two sides is $\sqrt{%
\frac{N-1}{N}}$. Moreover, we have just proven the length of the third side
is $\sqrt{2(1-\left| v_{j_{0}}\right| ^{2})}$. Since this is true regardless
of $S_{\psi }$, the projection of $V(\frac{1}{N}I,S_{\psi })$ onto $V(\frac{1%
}{N}I,E_{\psi })$ will be the same for all $S_{\psi }$. This means that all $%
S_{\psi }$ lie in the hyperplane, $C_{\psi }$, which is perpendicular to $V(%
\frac{1}{N}I,E_{\psi }).$ This hyperplane divides the set of states into two
regions: i) one which contains the plane and all states on the $\frac{1}{N}I$
side of the plane and ii) $G_{\psi },$ which is the open, connected set
which includes $E_{\psi }$ and all states on that side of $C_{\psi }$. A
state, $Q$, is in $G_{\psi }$ if and only if the projection of $V(\frac{1}{N}%
I,Q)$ onto $V(\frac{1}{N}I,E_{\psi }),$ is longer than the projection of $V(%
\frac{1}{N}I,S_{\psi })$ onto $V(\frac{1}{N}I,E_{\psi })$, i.e.$\left\langle
Q-\frac{1}{N}I,E_{\psi }-\frac{1}{N}I\right\rangle >\left\langle S_{\psi }-%
\frac{1}{N}I,E_{\psi }-\frac{1}{N}I\right\rangle =\left| v_{j_{0}}\right|
^{2}-\frac{1}{N}.$ Expanding the left hand side of this inequality and using
the fact that if $P$ is rank 1, then $\left\langle P,\frac{1}{N}%
I\right\rangle =Tr(\frac{1}{N}IP)=\frac{1}{N}$, we get $\left\langle
Q,E_{\psi }\right\rangle -\frac{1}{N}>\left| v_{j_{0}}\right| ^{2}-\frac{1}{N%
}.$

This proves the inequality in the theorem and it also shows why there are no
separable states in $G_{\psi }$. Indeed due to the convexity of the set of
separable states, if there were a separable state in $G_{\psi }$, then there
would have to be a pure, separable state in $G_{\psi }$. But this last
inequality shows that such a state would be closer to $E_{\psi }$ than is
possible by theorem (1).

The same reasoning can be applied to $F_{\psi }$. By (2) we see that the
projections (separable or entangled) which are furthest from $E_{\psi }$ are
those whose inner product with $E_{\psi }$ is 0. These are precisely the
ones which commute with $E_{\psi }.$ Since they all must lie on $F_{\psi }$,
it follows that there can be no states (separable or entangled) on the side
of $F_{\psi }$ that does not contain $E_{\psi }$. \ Hence all separable
states either lie on $C_{\psi }$ or $F_{\psi }$ or lie between them.

As for the last theorem, first note that the projection of $V(\frac{1}{N}%
I,S_{\psi })$ onto $V(\frac{1}{N}I,E_{\psi })$ has length $(\left|
v_{j_{0}}\right| ^{2}-\frac{1}{N})\sqrt{\frac{N}{N-1}}$ and the projection
of a furthest away pure state onto $V(\frac{1}{N}I,E_{\psi })$ has length $%
\frac{1}{N}\sqrt{\frac{N}{N-1}}$. Thus the distance between $C_{\psi }$ and $%
F_{\psi }$ is $\left| v_{j_{0}}\right| ^{2}.$ Since $\sum_{j=1}^{n_{1}}%
\left| v_{j}\right| ^{2}=1,$ the minimum $\left| v_{j_{0}}\right| $ is $%
\frac{1}{n_{1}}.$ To obtain the last theorem we express ${\Bbb C}^{N}$ as
the tensor product of ${\Bbb C}^{N_{1}}$ and ${\Bbb C}^{N_{2}}$ where $%
N_{1}=n_{\pi _{1}}\cdots n_{\pi _{k}}$ and $N_{2}=n_{\pi _{k+1}}\cdots
n_{\pi _{p}},$ with $\pi $ being the partition which makes $\frac{1}{f(\pi )}
$ maximum. Hence ${\Bbb C}^{N_{1}}={\Bbb C}^{n_{\pi _{1}}}\otimes \cdots
\otimes {\Bbb C}^{n_{\pi _{k}}}$ and ${\Bbb C}^{N_{2}}={\Bbb C}^{n_{\pi
_{k+1}}}\otimes \cdots \otimes {\Bbb C}^{n_{\pi _{p}}}.$ Any state which is
separable in ${\Bbb C}^{n_{1}}\otimes \cdots \otimes {\Bbb C}^{n_{p}}$ is
also separable in ${\Bbb C}^{N_{1}}\otimes {\Bbb C}^{N_{2}}$. However, these
latter states are all sandwiched between hyperplanes $C_{\psi }$ and $%
F_{\psi }$ which are associated with an entangled state for which $\left|
v_{j}\right| ^{2}=\frac{1}{N_{1}}$. It follows from what we said a moment
ago that for such a state the distance between $C_{\psi }$ and $F_{\psi }$
is $\frac{1}{N_{1}}\sqrt{\frac{N}{N-1}}=\kappa \frac{N}{N-1}.$

{\it Acknowledgement:}{\bf \ }This work was supported by a grant from the
Office of Naval Research.


\begin{references}
\bibitem{EPR}  A.Einstein, B.Podolsky, N.Rosen, Phys. Rev. {\bf 47} (1935)777

\bibitem{Bell}  J.S.Bell, Rev. Mod. Phys. {\bf 38} (1966) 447; and Physics 
{\bf 1} (1964) 195.

\bibitem{Bohm}  J.A.Wheeler and W.H.Zurek, editors, {\it Quantum Theory of
Measurement,} Princeton University Press, (1983)

\bibitem{Peres}  A.Peres, {\it Quantum Theory: Concepts and Methods}, Kluwer
Academic Press, (1995)

\bibitem{Ormes}  R.Omnes,{\it \ The Interpretation of Quantum Mechanics,}
Princeton University Press (1994); and {\it Understanding Quantum Mechanics}%
, Princeton University Press (1999)

\bibitem{popescu}  H.K.Lo, S.Popescu, and T.P.Spiller editors, {\it %
Introduction to Quantum Computing and Information, }World Scientific{\it \
(1998)}

\bibitem{williams}  C.Williams and S.Clearwater, {\it Explorations in
Quantum Computing,} Springer - Verlag (1998)

\bibitem{brassard}  C.H.Bennett, G.Brassard, C.Crepeau, R.Josza, A.Peres,
and W.K.Wooters, Phys. Rev. Lett. {\bf 70 (}1993) 1895

\bibitem{vitali}  D.Vitali, M.Fortunato and P.Tombesi, Phys.Rev.Lett. {\bf %
85 }(2000) 445

\bibitem{werner}  R.F.Werner, All Teleportation and Dense Coding Schemes,
quantph/0003070

\bibitem{shor}  P.Shor and J.Preskill, Phys.Rev.Lett {\bf 85 }(2000) 441

\bibitem{losalamos}  W.T.Buttler, {\it et.al.}, Phys.Rev.Lett. {\bf 81 (}%
1998) 3283

\bibitem{bennet}  C.H.Bennett, D.P.DiVincenzo, J.A.Smolin and W.KWooters,
Phys.Rev.A, {\bf 54} (1996) 3824

\bibitem{barnes}  D.A.Lidar, D.Bacon, J.Kempe and K.B.Whaley,
Decoherence-Free Subspaces for Multiple-Qubit Errors: (I) Characterisation,
e-print quant-ph/9908064v2. and Decoherence-Free Subspaces for
Multiple-Qubit Errors:(II) Universal, Fault-Tolerent Quantum Computation,
e-print quant-ph/0007013

\bibitem{note}  The references we give in this paragraph are offered as a
sampling. The literature on entanglement and its applications is large and
growing larger everyday. For instance, so far this year the LANL e-print
service has received over 100 e-prints with the word 'entanglement' in the
title. The number of works that deal with entanglement, obviously, is much
larger.

\bibitem{pit}  A.O.Pittenger and M.H.Rubin, Note on Separability of the
Werner states in arbitrary dimensions, quant-ph/0001110v2

\bibitem{braun}  S.L. Braunstein, {\it et al, }Phys.Rev.Lett, {\bf 83 (}%
1999) 1054

\bibitem{caves}  C.M. Caves and G.J. Milburn, Qutrit Entanglement,
quant-ph/9910001

\bibitem{dur}  W.Duer, J.I. Cirac, and R. Tarrach, Phys.Rev.Lett. {\bf 83 }%
(1999) 3562
\end{references}
\end{document}